\documentstyle[12pt]{article}
\begin{document}
\thispagestyle{empty}
\begin{center}
\LARGE \tt \bf{Double cosmic walls in Teleparallel Gravity}
\end{center}
\vspace{1.0cm}
\begin{center} {\large L.C. Garcia de Andrade\footnote{Departamento de
F\'{\i}sica Te\'{o}rica - Instituto de F\'{\i}sica - UERJ
Rua S\~{a}o Fco. Xavier 524, Rio de Janeiro, RJ
Maracan\~{a}, CEP:20550-003 , Brasil.
e-mail.:garcia@dft.if.uerj.br}}
\end{center}
\vspace{1.0cm}
\begin{abstract}
An example is given of a plane topological torsion defect representing a cosmic wall double wall
in teleparallel gravity.The parallel planar walls undergone a repulsive gravitational force due 
to Cartan torsion.This is the first example of a non-Riemannian double cosmic wall.It is shown 
that the walls oscillate with a speed that depends on torsion and on the surface density of the 
wall.Cartan torsion acts also as a damping force reducing the speed of oscillation when it is stronger.  
\end{abstract}
\vspace{1.0cm}
\begin{center}
\end{center}
\newpage
\paragraph*{}
A solution of Einstein equations representing parallel planar topological defects  \cite{1} have been investigated in detail by Letelier \cite{2}.Although Riemannian topological defects are defined as the spacetime manifolds where the Riemann-Christoffel curvature tensor vanishes only off the defects, we here define teleparallel defects as solutions of the teleparallel condition  
\begin{equation}
R^{i}_{jkl}(U_{4})=0
\label{1}
\end{equation}
and have the same form of the metric of the Riemannian defect.In this Letter we shall consider the planar parallel cosmic walls in the context of the teleparallel theory of gravity as was developed by Einstein himself during the period in 1928 in Berlin \cite{3}.Let us consider the double cosmic wall solution as \cite{2}
\begin{equation}
ds^{2}=({\omega}^{0})^{2}-({\omega}^{1})^{2}-({\omega}^{2})^{2}-({\omega}^{3})^{2}
\label{2}
\end{equation}
where the basis 1-forms $ {\omega}^{i} $ (i=0,1,2,3) are given by
\begin{equation}
\begin{array}{llll}
{\omega}^{0}= e ^{-2{\pi}{\sigma}(h^{2}-z^{2})}dt \nonumber \\
{\omega}^{1}=2z e ^{-2{\pi}{\sigma}(h^{2}-z^{2})}dz \nonumber \\
{\omega}^{2}=e ^{-2{\pi}{\sigma}(h^{2}-z^{2}-t)}dx\\
{\omega}^{3}=e ^{-2{\pi}{\sigma}(h^{2}-z^{2}-t)}dy \nonumber \\
\end{array}
\label{3}
\end{equation}
where ${\sigma}$ is the surface matter density of the wall and $h$ is the separation between the walls.Here since we shall consider that the walls may move relative to each other this quantity will depend of the coordinates t and z.In the case of Letelier double Riemannian cosmic wall this amount is fixed.The torsion 1-forms are chosen such that the connection 1-forms ${\omega}^{i}_{j}$ vanish.Thus from the second Cartan equation of differencial exterior forms 
\begin{equation}
{{R}^{i}}_{k} = d{{\omega}^{i}}_{k} + {{\omega}^{i}}_{l} \wedge {{\omega}^{l}}_{k}
\label{4}
\end{equation}
where the curvature 2-form is $R^{i}_{j}=R^{i}_{jkl}{\omega}^{k}\wedge{\omega}^{l}$, we see that the teleparallel condition (\ref{1}) is fulfilled.From the first Cartan equation 
\begin{equation}
T^{i}=d{\omega}^{i}+{{\omega}^{i}}_{k} \wedge {\omega}^{k}
\label{5}
\end{equation}
yields
\begin{equation}
\begin{array}{llll}
T^{0}= 4{\pi}{\sigma}(hh_{z}-z)e^{-2{\pi}{\sigma}(h^{2}-z^{2})}dz\wedge dt\nonumber \\
\\
T^{1}= -4{\pi}{\sigma}zh\dot{h}e^{-2{\pi}{\sigma}(h^{2}-z^{2})}dt \wedge dz \nonumber  \\
\\
T^{2}=-4{\pi}{\sigma}[2h\dot{h}-1]e^{-2{\pi}{\sigma}(h^{2}-z^{2}-t)}[[2h\dot{h}-1]dt\wedge dx+[hh_{z}-z]dz\wedge dx \ \nonumber \\
\\
T^{3} =-4{\pi}{\sigma}e^{-2{\pi}{\sigma}(h^{2}-z^{2}-t)}[[2h\dot{h}-1]dt\wedge dy+[hh_{z}-z]dz\wedge dy \nonumber \\
\end{array}
\label{6}
\end{equation}
Let us now consider to simplify matters that torsion components $T^{i}_{jk}$ in the torsion 2-forms 
\begin{equation}
T^{i}=T^{i}_{kl}{\omega}^{k} \wedge {\omega}^{l}
\label{7}
\end{equation}
yields the following partial differential equations
\begin{equation}
\begin{array}{llll}
(hh_{z}-z)e^{-2{\pi}{\sigma}(h^{2}-z^{2})}=T^{t}_{zt}=K_{1} \nonumber \\
\\
z(h\dot{h}-1)e^{-2{\pi}{\sigma}(h^{2}-z^{2})}=T^{z}_{tz}=K_{2} \nonumber \\
\\
T^{x}_{tx}=(h\dot{h}-1)e^{-2{\pi}{\sigma}(h^{2}-z^{2}-t)} \nonumber \\
\\
T^{y}_{ty}=(hh_{z}-z)e^{-2{\pi}{\sigma}(h^{2}-z^{2}-t)} \nonumber\\
\end{array}
\label{8}
\end{equation}
where we have chosen the first two components above as the constants $K_{1}$ and $K_{2}$ respectively.The first two equations are
\begin{equation}
(hh_{z}-z)e^{-2{\pi}{\sigma}(h^{2}-z^{2})}=K_{1}
\label{9}
\end{equation}
and
\begin{equation}
z(h\dot{h}-1)e^{-2{\pi}{\sigma}(h^{2}-z^{2})}=K_{2}
\label{10}
\end{equation}
The first equation yields
\begin{equation}
h^{2}=z^{2}+2K_{1}z-\frac{4{\pi}{\sigma}z^{3}}{3}+f(t)
\label{11}
\end{equation}
where $f(t)$ is in principle an arbitrary integration function of time to be determined later.Integration of the secont PDE yields
\begin{equation}
h^{2}=t[1+K_{2}(1-2{\pi}{\sigma}z^{2})]
\label{12}
\end{equation}
where we have used the approximation $h^{2}<<<z^{2}$.By equating the equations (\ref{11}) and (\ref{12}) allow us to determine the function $f(t)$ as  
\begin{equation}
f(t)=t[1+K_{2}(1-2{\pi}{\sigma}z^{2})]-z^{2}-2K_{1}z(1-\frac{2{\pi}{\sigma}z^{2}}{3})
\label{13}
\end{equation}
substitution of equation (\ref{13}) into equation (\ref{11}) one yields
\begin{equation}
h=[t(1+K_{2}(1-2{\pi}{\sigma}z^{2}))-\frac{4{\pi}{\sigma}z^{3}}{3})]^{\frac{1}{2}}
\label{14}
\end{equation}
Performing the second derivatives of expression (\ref{14}) with respect to the time and z-coordinates respectively yields
\begin{equation}
-\frac{{\partial}^{2}h}{{\partial}t^{2}}=\frac{1}{4}{\phi}^{-{\frac{3}{2}}}[1+K_{2}(1-2{\pi}{\sigma}z^{2})]^{2}
\label{15}
\end{equation}
and 
\begin{equation}
\frac{{\partial}^{2}h}{{\partial}z^{2}}={\pi}^{2}{\sigma}^{2}{\phi}^{-\frac{3}{2}}[K_{2}t-K_{1}z]
\label{16}
\end{equation}
where \begin{equation}
{\phi}=[t(1+K_{2}(1-2{\pi}{\sigma}z^{2})-\frac{4{\pi}{\sigma}z^{3}}{3})]
\label{17}
\end{equation}
an arrangement of these last two equations yields the wave equation
\begin{equation}
\frac{{\partial}^{2}h}{{\partial}z^{2}}-{\frac{1}{v^{2}}}\frac{{\partial}^{2}h}{{\partial}t^{2}}=0
\label{18}
\end{equation}
where
\begin{equation} v^{2}=[\frac{1+K_{2}(1-2{\pi}{\sigma}z^{2})}{{\pi}{\sigma}(K_{2}t-K_{1}z)}]^{2}
\label{19}
\end{equation}
From this expression we see that the walls oscillate under the influence of torsion.A particular interesing case is when torsion component $K_{1}$ vanishes 
\begin{equation}
v=\frac{1}{{\pi}{\sigma}}[\frac{1}{K_{2}t}+(1-2{\pi}{\sigma}z^{2})]
\label{20}
\end{equation}
physically this equation means that when time goes to infinity and z approaches zero,or the 
walls approach each other the speed of oscillation is reduced.Besides when torsion grows the 
speed of oscillation also goes to zero and torsion produces a damping effect on the oscillation 
of both the walls.This effect certainly would produce gravitational waves and even torsion waves.
A more detailed investigation of the reaction of the gravitation radiation on these teleparallel 
walls may appear elsewhere.Other types of torsion defects \cite{4,5} producing radiation can be 
considered in near future.
\section*{Acknowledgments}
\paragraph*
I would like to express my gratitude to Prof.P.S.Letelier and Prof.A.Wang for helpful discussions on the subject of this paper. Financial support from UERJ and CNPq. is gratefully acknowledged.


\begin{thebibliography}{5}
\bibitem{1}A.Vilenkin and P.Shellard,Cosmic Strings and other Topological Defects,(1995),Cambridge university Press.
\bibitem{2}P.S.Letelier,(1990),Class. and Quantum Gravity.7,L 203.
\bibitem{3}A.Einstein,(1928)Berliner Sitzungsber.217.
\bibitem{4}L.C.Garcia de Andrade,(1999)Class. and Quantum Grav.16,6,2097.
\bibitem{5}L.C.Garcia de Andrade,Phys.Lett.A,(1999)256.
\end{thebibliography}
\end{document}